\documentclass[aps,prl,twocolumn,groupedaddress,showpacs]{revtex4}

\usepackage{graphicx}

\begin{document}

\title{Giant magnetic enhancement in Fe/Pd films and its influence on the 
magnetic interlayer coupling}

\author{Erik Holmstr\"{o}m}
\email[]{erik.holmstrom@fysik.uu.se}

\affiliation{Condensed Matter Theory Group, Physics 
Department Uppsala University}
\author{ Lars Nordstr\"{o}m }
\affiliation{Condensed Matter Theory Group, Physics 
Department Uppsala University}
\author{A.M.N.\ Niklasson}
\affiliation{Theory Division, Los Alamos National Laboratory, 
Los Alamos, NM 875 45, USA}

\date{\today}

\begin{abstract}
The magnetic properties of thin Pd fcc(001) films with embedded
monolayers of Fe are investigated by means of first principles
density functional theory. 
The induced spin polarization in Pd
is calculated and analyzed in terms of quantum interference within the
Fe/Pd/Fe bilayer system.
An investigation of the magnetic enhancement effects on the spin polarization
is carried out and its consequences for the magnetic interlayer coupling are discussed. 
In contrast to {\it e.g.} the Co/Cu fcc(001) system we find a large
effect on the magnetic interlayer coupling due to magnetic enhancement
in the spacer material. In the case of a single embedded Fe monolayer we find an
induced Pd magnetization decaying with distance $n$ from the magnetic layer as 
~$n^{-\alpha}$ with $\alpha \approx 2.4$. 
For the bilayer system we find a giant magnetic enhancement (GME) that 
oscillates strongly due to interference effects. 
This results 
in a strongly modified magnetic interlayer coupling, both in phase and magnitude, 
which may not be described in the pure  Ruderman-Kittel-Kasuya-Yoshida 
(RKKY) picture. 
No anti-ferromagnetic coupling was found and by comparison with
magnetically constrained calculations we show that the overall ferromagnetic 
coupling can be
understood from the strong polarization of the Pd spacer.
\end{abstract}


\maketitle

\section{Introduction}
In order to understand itinerant magnetism, paramagnetic
materials with a high spin susceptibility are of special
interest. These materials are on the border of being
ferromagnetic and small perturbations may therefore
lead to a spontaneous spin polarization. A reduction
of the dimension may, for example, induce a magnetic moment
in thin Pd films placed on Ag fcc(001) as a result of
the finite size effects in the Pd film \cite{Niklasson97}. 
Another example is the large polarization cloud around Fe impurities in Pd. 
The cloud has a radius of about 10 {\AA} and a total 
magnetic moment of 12 $\mu_{B}$ \cite{Low66}. The moment 
of the Fe impurity itself is on the order of 3 $\mu_{B}$ 
which means that about 9 $\mu_{B}$ 
is coming from the induced moments in Pd.
The effect can theoretically be understood in
terms of enhancement of the paramagnetic Pauli spin susceptibility
that strongly modifies the magnetic properties of the unperturbed
paramagnetic ground state. In this paper we investigate
the magnetization of Pd when the Fe impurity is
substituted by a two-dimensional layer.
We also analyze the magnetic interaction, i.e. the
magnetic interlayer coupling (MIC) between two separate
monolayers of Fe in a Pd host. 
This could lead to
large moments of the embedded Pd host. In that case
the electronic structure of Pd may be strongly modified
and the Ruderman, Kittel, Kasuya, Yoshida (RKKY) or quantum-well (QW)
like theories describing the magnetic interlayer coupling that are based
on the properties of the unperturbed paramagnetic spacer bulk material 
may break down, c.f. the Fe/Cr/Fe system \cite{Niklasson99}.

The MIC has attracted a lot of interest since its discovery 
by Gr\"unberg {\it et al.} in 1986 \cite{Grunberg86}. 
It is a long range interaction between 
magnetic layers through paramagnetic spacer material and combined with the 
giant magnetoresistance effect \cite{Baibich88} it is one of the main
mechanisms in magnetic sensor and memory devices.
The MIC is theoretically well understood
\cite{Stiles93,Stiles99,Bruno95,Bruno97} for the case of a paramagnetic, insulating
or disordered alloy spacer and it has been shown to be well described
by the RKKY \cite{Ruderman_kittel54,Kasuya56,Yoshida57} coupling
originally explaining the interaction between local magnetic impurities 
in a paramagnetic host. A related way of describing 
the MIC is the QW picture \cite{Niklasson99JPCM,Mathon92,NiklassonPRB96,Bruno99}, 
where the coupling is viewed in terms of 
confined states in the spacer that have different energies depending on 
spacer thickness and orientation between the magnetic layers.

In the last ten years the  Fe/Pd multilayers has drawn a lot of attention, 
mostly because the first measurements indicated a weak 
anti-ferromagnetic (AFM) coupling for spacer thicknesses around 12 monolayers 
(ML) in a 
Fe$_{5.7}$/Pd$_{x}$/Fe$_{9.6}$ (100)
system (Celinski {\it et al.} \cite{Celinski91}) that looked very promising 
for applications. 
Later Childress {\it et al.} \cite{Childress94} investigated 
the same system and found no trace of an AFM coupling in the thickness range 
from $\sim$ 5
to 25 ML. They also noted that a weak in-plane uniaxial anisotropy that changes 
with spacer 
thickness may be mistaken for an AFM coupling. However, in the latest experiment 
on the 
MIC, Lucic {\it et al.} found a strong AFM coupling for a Pd thickness of 6 ML 
\cite{Lucic99}. Lucic also showed that the interface morphology plays an important 
role and 
that the AFM peaks are only present in samples with very smooth interfaces 
($\sim$ 1000 nm 
size terraces). The switching from ferromagnetic (FM) to AFM coupling also 
agrees with the calculated results by Stoeffler {\it et al.} for one of 
their considered 
structures of the Pd spacer, the constant atomic volume 
structure (CAV). In this structure the Pd maintains its bulk volume but adapts the 
interplanar distance in order to fit the Fe bcc(100) interface \cite{Stoeffler94}.  The results
by Stoeffler {\it et al.} are for spacer thicknesses up to 6 ML so the width of this AFM
region cannot be determined. The results up to now on the MIC, both experimental 
and theoretical have been on multilayers with thick bcc (100) Fe layers sandwiched with 
fcc (100) Pd.
Recently Cros {\it et al.} found a high spin Fe phase in Fe/Pd(100) when the Fe layers 
are very thin (1 to 3 ML). They also found that the Fe layers adapt to the Pd fcc structure with 
almost the same lattice constant \cite{Cros00}. 

This paper is outlined as follows. Firstly, 
we present some calculational details, where we describe
the linear muffin-tin
orbitals (LMTO) Green's function technique used in the density functional
calculations, then we describe how effects due to multiple scattering,
{\it i.e.} interface interference terms, can be extracted and analyzed separately, and how
magnetic enhancement effects can be estimated by a comparison
with magnetically quenched calculations. Thereafter
we show the result of the magnetization induced in Pd from
a planar interface perturbation that is a single monolayer of Fe fcc (001)
embedded in the Pd fcc host. In the next section we calculate
the induced magnetization of the bilayer system, with two separate Fe monolayers
embedded in the Pd host, and we estimate
the effect due to interface interference.  Subsequent to that, 
the magnetization results are discussed in terms of quantum well states.
At the end we present the calculated MIC and by a comparison with magnetically
constrained calculations and the MIC of the Co/Cu
system we discuss the possibility to apply RKKY theory for
describing the magnetic interlayer coupling in magnetically
enhanced multilayers.

\section{Calculational details}
\subsection{Total energy calculations}
The {\it ab initio} density functional calculations were performed within
the local spin density approximation applying an interface
Green's function technique developed by Skriver and Rosengaard
\cite{Skriver91}. The method is based on the LMTO method \cite{Andersen75,Skriver84} within the
tight-binding \cite{Andersen85}, frozen core, and atomic-sphere
approximations. The local spin density
approximation as parameterized by Vosko, Wilk, and Nusair
\cite{Vosko80} was used. An advantage of the Green's function technique is
that it ensures a correct description of the loss of translational
symmetry perpendicular to the interface, without the use of an
artificial slab or supercell geometry.
The studied bilayers consist of two layers of Fe with a thickness 
of 1 ML embedded in bulk Pd. 
The Fe layers are separated by $N$ layers of Pd, where 
$1\leq N \leq 16$.
All calculations were performed in the Pd bulk fcc lattice with a lattice constant 
a=3.89 \AA, neglecting structure relaxations. The 
direction of growth was (100). This particular choice of structure
is justified by the results by Cros {\it et al} \cite{Cros00}.

The MIC was calculated as the difference 
in total energy between the situation of  AFM
aligned Fe layers and the system with  FM aligned 
Fe layers

\begin{equation}
J_{MIC}(N)=E^{tot}_{AFM}(N)-E^{tot}_{FM}(N).
\end{equation}

\noindent
Furthermore, we found that 
528 special k-points were needed in the irreducible part of the 
2-dimensional Brillouin zone to obtain k-point convergence of the MIC.

\subsection{Multiple scattering contribution to the magnetic moment} 

To analyze the multiple scattering contribution to the magnetic 
moment in the spacer the method by Niklasson {\it et al.} described in detail in ref. 
\cite{NiklassonPRB96} was used. The main idea is that the perturbed Green's
function of the embedded Pd film G$^{\sigma}$ can be expressed in terms
of the corresponding unperturbed paramagnetic bulk Green's function $G_0$ as
\begin{equation}\label{TotGreenexp}
G^{\sigma} = G_{0}  + \Delta G_{L}^{\sigma} + 
\Delta G_{R}^{\sigma} +  \Delta G_{QW}^{\sigma} .
\end{equation}
Here $\Delta G_{L}^{\sigma}$ and $\Delta G_{R}^{\sigma}$ are the independent
planar perturbations arising from the two non-interacting Fe monolayers, and
$\Delta G_{QW}^{\sigma}$ is the term due to interference in the QW
like structure formed by the Fe/Pd/Fe system. By subtracting a superposition of
the independent magnetic interface perturbations, calculated from a single Fe ML 
embedded in the Pd host, from the magnetization profile of the fully interacting
Fe/Pd/Fe bilayer system the contribution to the magnetic moment from the interference 
term can be estimated.  
In this way the effects of quantum interference on the polarization 
of the spacer 
due to the occurrence of confined, so called QW states, can be
separated and analyzed.

\subsection{Magnetic enhancement}
When a material is exposed to a magnetic field its 
magnetization is determined by the magnetic susceptibility. 
However, the magnetization of the material induces an internal magnetic 
field that influences the neighbouring atoms and the resulting magnetization 
is determined by the response to the sum of the applied field 
and the induced field inside the material. This ``self-consistency effect'' is 
referred to as exchange 
enhancement and is large in elements with a high 
susceptibility such as Palladium. 
The enhanced susceptibility 
can be estimated from the un-enhanced Pauli spin susceptibility, $\chi_{0}$,
by

\begin{equation}\label{eq:susc}
\chi = \frac{\chi_{0}}{1-I\chi_{0}}
\end{equation}

\noindent
where $I$ is the Stoner exchange parameter that specifies the 
intra-atomic magnetic coupling in the material.
The exchange enhancement, that may lead to a complete change
of the electronic structure, is hard to estimate and is generally
not included in, for example, the RKKY theory describing the
magnetic interlayer coupling in magnetic multilayers. It is thus
of great interest to try to understand the magnitude of this effect.

Methods of obtaining the MIC that use the force theorem or 
frozen potentials for the spacer
atoms are based on the RKKY description of the MIC and although
very successful for calculating asymptotic behaviour for many 
multilayer systems, they do not take the enhanced susceptibility 
into account. 
For technical reasons, we cannot employ this type of approximations 
directly in our computer program and
in order to mimic such an approach we have 
imposed a  constraint $\chi=\chi_{0}$, {\it i.e.} $I=0$,
that we from now refer to as the quenched case. 

This limit can be achieved by imposing the condition that 
the exchange correlation potential for some, 
or all Pd atoms in the system only contains the paramagnetic 
part. It is important to notice that since the magnetic moments are 
determined by the Green's function for the whole sample there will 
still be nonzero magnetic moments on the quenched Pd atoms.
The Pd magnetization
is quenched in order to critically test whether  a perturbation 
around a paramagnetic ground state is valid, as in a RKKY model.

\section{Results}

\subsection{ Interface Induced Magnetization}

In Table \ref{tab:mprofile} we show the calculated magnetic profile of 
one embedded Fe layer in Pd in the case of enhanced and quenched Pd 
atoms. The profile corresponds to a calculation of the magnetic moments 
from the $\Delta G_{L}^{\sigma}$ terms in eq.({\ref{TotGreenexp}}).

 In the enhanced situation one can see that there is a weak
oscillation of the Pd moments around a positive value and a 
decay with the distance $n$ monolayers from the magnetic layer. The 
findings are in agreement with results for Fe impurities in Pd, 
both experimental \cite{Low66} and theoretical \cite{Zeller93}.
By a least square fit of the calculated enhanced profile to the approximate
decaying sinusoidal description 
$\sim n^{ - \alpha}( C\sin(\beta n + \Phi) +\xi)$
we find that $\alpha \approx 2.4$, $C \approx 0.63$, 
$\beta \approx 1.11$,  $\Phi \approx 3.5$ 
and $\xi \approx 0.96$. In the quenched situation, the positive
bias has disappeared and the magnetic profile oscillates around the 
zero level but the interface Pd atom still maintains most of its polarization. 
Worth pointing out here is that the decay is 
here obtained from a fit to all 
moments including the interface 
Pd atom and does not necessarily describe the exact asymptotic result.

\begin{table}
\begin{center}
  \begin{tabular}{l *{2}{r@{.}l} | l   *{2}{r@{.}l}}
    \hline\hline
    \multicolumn{1}{l} { Layer}  &  \multicolumn{2}{c} {  Enhanced} & 
     \multicolumn{2}{c|}{  Quenched} & \multicolumn{1}{|l} {Layer}  &  
      \multicolumn{2}{c} {  Enhanced} & \multicolumn{2}{c}{  Quenched}\\
    \hline
     $Fe$               &\ \ 3&0378     &    \    3&0512  &     $Pd_{10}$     &\ \ 0&0026     &   \    -0&0005\\
     $Pd_{1}$      &\ \ 0&3330     &    \    0&2251  &     $Pd_{11}$     &\ \ 0&0005     &   \     0&0012\\
     $Pd_{2}$      &\ \ 0&1396     &    \    0&0412  &     $Pd_{12}$     &\ \ 0&0006     &   \     0&0015\\
     $Pd_{3}$      &\ \ 0&0978     &    \    0&0048  &     $Pd_{13}$     &\ \ 0&0021     &   \     0&0005\\
     $Pd_{4}$      &\ \ 0&0561     &    \   -0&0029  &     $Pd_{14}$     &\ \ 0&0029     &   \    -0&0007\\
     $Pd_{5}$      &\ \ 0&0236     &    \   -0&0002  &     $Pd_{15}$     &\ \ 0&0018     &   \    -0&0008\\
     $Pd_{6}$      &\ \ 0&0079     &    \    0&0026  &     $Pd_{16}$     &\ \ 0&0004     &   \     0&0001\\
     $Pd_{7}$      &\ \ 0&0053     &    \    0&0004  &     $Pd_{17}$     &\ \ 0&0000     &   \     0&0008\\
     $Pd_{8}$      &\ \ 0&0067     &    \   -0&0018  &     $Pd_{18}$     &\ \ 0&0005     &   \     0&0006\\
     $Pd_{9}$      &\ \ 0&0053     &    \   -0&0021  &     $Pd_{19}$     &\ \ 0&0009     &   \    -0&0001\\
    \hline\hline
  \end{tabular}
\caption[]{The magnetic profile of the 
           Pd$_{\infty}$/Fe$_{1}$/Pd$_{\infty}$ system as calculated 
           with unconstrained susceptibility (Enhanced) and
           constrained to the un-enhanced Pauli susceptibility 
           on the Pd atoms (Quenched). The unit is $\mu_{B}$/atom.}
\label{tab:mprofile}
\end{center}
\end{table}

\begin{figure}
    \includegraphics*[angle=0,width=0.45\textwidth]{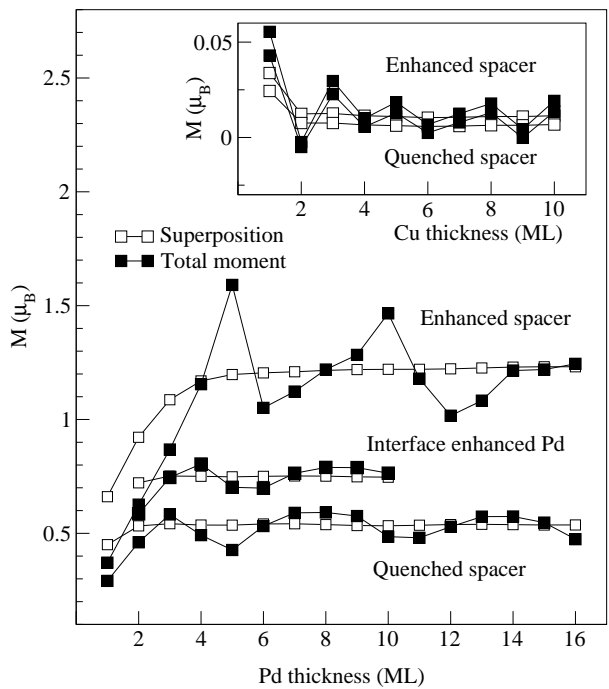}
    \caption{The total magnetic moment of the spacer as calculated by a 
                    superposition of magnetic profiles and by doing a self 
                    consistent calculation of the system, both for quenched 
                    and enhanced Pd spacer layers. The difference between the 
                    curves calculated by a superposition indicates a giant magnetic 
                    enhancement (GME) effect of about 
                    0.7 $\mu_{B}$ arising from the non interacting Fe layers. By comparing 
                    total moments for the enhanced and quenched calculations
                    the interference contribution to the GME can be estimated. The inset
                    shows the same calculation for the Co/Cu system.
    \label{fig:Mbidrag}}
\end{figure}

Moving on to the case of two embedded Fe layers we see in 
Fig.(\ref{fig:Mbidrag}) the total magnetic moment of the spacer 
together with the superposition of magnetic profiles from a single Fe 
monolayer in Pd. The figure shows the results from calculations where
all Pd atoms are quenched, when the interface Pd atoms are left enhanced 
and when all Pd atoms are enhanced. 
In all cases the superimposed curves reach a finite, positive value, the 
completely quenched
curve about $0.5\mu_{B}$, the interface enhanced curve $0.75\mu{B}$ 
and the enhanced curve about 
$1.2\mu{_B}$. 
The constant value is reached 
faster in the completely quenched case, already at 2 ML spacer thickness compared 
to the enhanced case where the constant value is reached at about 5 ML.
This effect is natural since the polarization of Pd in the case of one embedded 
Fe ML is stronger and more long range in the enhanced case compared to the 
quenched case (see Table \ref{tab:mprofile}).
The shift between the enhanced and completely quenched superimposed 
curves is about 
$0.7\mu_{B}$ which must then be attributed to enhancement in Pd in
the absence of multiple interface scattering effects. This is a giant magnetic
enhancement (GME) effect which is completely absent in the case of the corresponding
Co/Cu system where the enhanced and quenched moments are virtually on top
of each other as seen in the inset.

The additional effect of multiple scattering, i.e. due to interference between the
two Fe monolayers, is seen from the difference between the superimposed
curves and the full calculation of the total moments in Fig.(\ref{fig:Mbidrag}). 
This difference corresponds to the magnetic moments calculated from the 
$\Delta G_{QW}^{\sigma}$ terms in eq.({\ref{TotGreenexp}}).
In both situations the magnetic moment oscillates when the multiple 
scattering contribution is added with a period 
 that is compatible with 
the Fermi surface nesting vector of Pd ( 5.7 ML) \cite{Niklasson97,Stiles93}. 
By inspecting the three cases in 
Fig.(\ref{fig:Mbidrag}) one can see that the enhancement introduces a 
phase shift of the magnetization oscillation. At the same time the enhancement 
changes the character of the curve to be more sawtooth shaped. Note that the 
sawtooth shape does not appear in the intermediate case, with interface enhanced Pd, 
where the band matching 
at the interfaces is very similar to the enhanced case.

\noindent

\subsection{Quantum well states}

\begin{figure*}
   \includegraphics*[angle=0,width=0.45\textwidth]{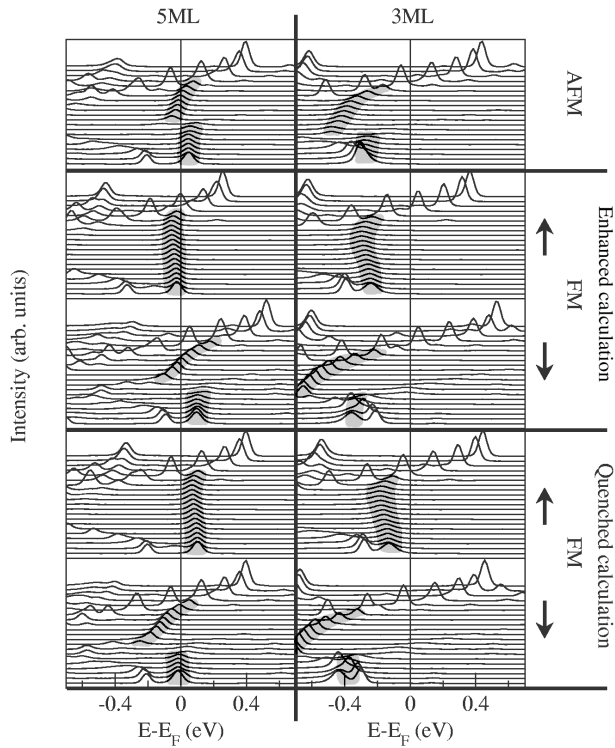}
    \caption[ ]{ The $\overline{\rm k}$-resolved spectral density 
                 of a 5 ML and 3 ML system for the 
                 FM and AFM configuration. The lower block is the quenched
                 and the middle block the enhanced calculation.
                 Each part of the figure contains several curves
                 where the lowest one is the spectral density at the 
                 $\overline{\Gamma}$
                 point. The other curves are also the spectral density but at k-points 
                 shifted in the $\overline{\rm M}$-direction.
                 The curves are individually shifted in the y-direction in order to be 
                 plotted in the same graph. The QW states can be seen as 
                 distinct peaks with 
                 a small dispersion and have been shaded for clarity. The AFM result 
                 is very similar for the quenched and enhanced calculation and only
                 the quenched result is shown here. The arrows indicate the different
                 spin channels. The thin
                 vertical lines indicate the Fermi energy in all graphs. 
                 \label{fig:QWstates}}
\end{figure*}

The big effect on the magnetization from multiple scattering is an indication
of QW states in the spacer as was shown by Niklasson {\it et al} 
\cite{NiklassonPRB96}.  In order to
 investigate this possibility we have calculated 
the spectral density along 
the line $\overline{\Gamma}-\overline{\rm M}$ in the 2 dimensional
Brillouin zone for 
all different configurations and spacer thicknesses as illustrated in 
Fig.(\ref{fig:QWstates}) for some cases.
The QW states can be identified by the change in energy as function of spacer 
thickness and we find very pronounced states in this system.
The first thing we see by looking at all states for all spacer thicknesses is 
the counter intuitive effect that the 
aliasing moves the states up in energy with 
increasing spacer thickness.
Secondly, the QW states in the spin-up channel has almost no 
dispersion at all whereas the spin-down state is divided into two disperse parts due 
to weak hybridization with Fe at the interfaces. We interpret this as an indication that 
the spin up state will influence the system to a greater extent than the spin down 
state.

For the AFM case, only the spectral density 
for the quenched calculation is shown. The reason is that we have not found any 
AFM case where the positions of the QW states differ between the enhanced and 
quenched calculations. The similarity is probably due to the fact that the total 
magnetic moment of the Pd spacer in the AFM system is always constrained to be 
zero by symmetry. 

In the quenched calculations of the FM systems we always find spin split QW 
states that are shifted in the opposite direction
to the bands {\it i.e.} the spin down state is shifted down 
in energy as compared to the spin up state. This can be explained by the different
boundary conditions at the interfaces for the two spin channels.  

In the enhanced FM systems the biggest effect on the QW state is the 
$0.7\mu_{B}$ polarization that was found in Fig.(\ref{fig:Mbidrag}) to 
be caused by the Fe layers when the interlayer interaction is neglected. 
The polarization is the same for all 
thicknesses above 5 ML resulting in a shift towards lower energies of the spin up QW states.
When the states are shifted they will pass the Fermi energy at a larger spacer thickness 
compared to the quenched case, resulting in a phase shift as can be seen by comparing 
Fig.(\ref{fig:QWstates})  and Fig.(\ref{fig:Mbidrag}) for the 5 ML case. For this spacer 
thickness, the spin up state has not passed the Fermi energy in the enhanced case but is 
just above in the quenched case.

The sawtooth shape of the magnetization in Fig.(\ref{fig:Mbidrag}) looks very similar
to the change in integrated density of states between the continuous and quantized 
situations in the article by Bruno \cite{Bruno95} Fig.(2).
There the sawtooth shape is originating from large reflection coefficients at the interfaces.
Coefficients close to one gives the sawtooth shape whereas smaller values result in a 
sinusoidal shape similar to our results for the completely quenched case.
An equivalent way of explaining the two shapes in Brunos article is that with complete 
reflection the QW state in the spacer will contribute like a delta function to the DOS 
of the spacer. When the spacer thickness is changed, the entire QW state will pass
the Fermi energy and give a step contribution to the integrated DOS which means a 
sawtooth shaped change in integrated DOS. If the reflection coefficients decrease,
the delta function will be smeared out in energy. Hence, the step in integrated DOS 
will be smeared and the change in integrated DOS will be smoothed towards a 
sinusoidal shape.

In our case we do not expect the reflection coefficients to change when the 
enhancement in introduced. This is confirmed both by the interface enhanced calculation 
in Fig.(\ref{fig:Mbidrag}) where the shape of the magnetization curve is sinusoidal
but the band matching at the Fe/Pd interface is very similar to the fully enhanced case,
and also by the similar shapes of the QW states in Fig.(\ref{fig:QWstates}). 
From these observations we
conclude that the system uses the extra freedom to split the bands in order to 
avoid the smeared QW state at the Fermi energy. The result is 
effectively the same as if the reflection coefficients at the interfaces 
were close to unity.

\noindent 
\begin{figure}
    \includegraphics*[angle=0,width=0.45\textwidth]{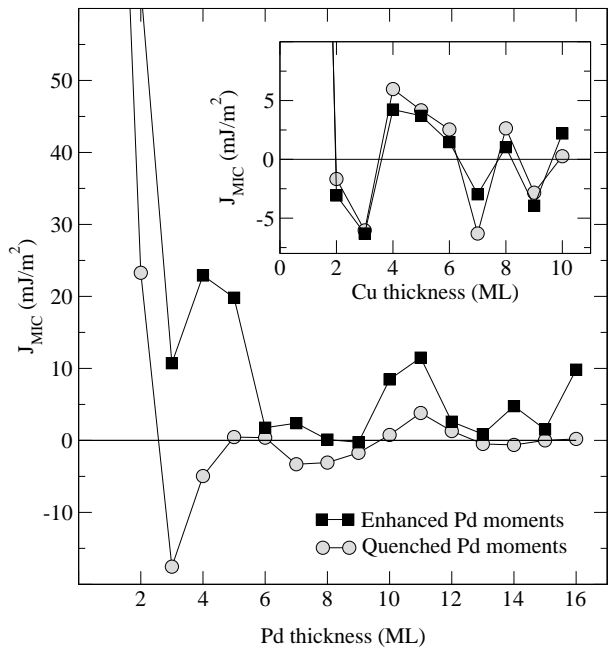}
    \caption[]{The MIC of the 
               Pd$_{\infty}$/Fe$_{1}$/Pd$_{N}$/Fe$_{1}$/Pd$_{\infty}$ 
               system for the quenched and enhanced calculations. The inset shows the 
               results for the corresponding Co/Cu system. \label{fig:MIC}}
\end{figure}

\subsection{MIC and enhancement}
Because of the GME effect in the system
arising from each Fe interface separately and the additional magnetization from the 
interface interference we may expect
a very large effect on the magnetic interlayer coupling. The MIC of the 
Pd$_{\infty}$/Fe$_{1}$/Pd$_{N}$/Fe$_{1}$/Pd$_{\infty}$ system is shown in
 Fig.(\ref{fig:MIC}). The two curves are the results 
of the enhanced  total energy calculation and the quenched 
calculation.
Corresponding calculations for the
Cu$_{\infty}$/Co$_{1}$/Cu$_{N}$/Co$_{1}$/Cu$_{\infty}$ system is
shown in the inset for comparison. The Co/Cu results are consistent with earlier calculations
\cite{Lang93}.

Both the two MIC results, with quenched Pd moments and enhanced moments show 
the same period between 5 and 6 ML as would be expected from RKKY theory in the limit of 
infinite spacer thickness.  
The largest difference between the two cases is that 
the quenched result
oscillates around zero comparably to the Co/Cu system and does 
not show the FM bias that
is present in the enhanced calculation. 
Also the decays with increasing spacer thickness are 
very different. By performing a fit to the decaying function
$ n^{ - \alpha}( C\sin(\beta n + \Phi) +\xi)$ for values $n>3$
we obtain $\alpha \approx 0.9$, $C \approx 45$, 
$\beta \approx 1.03$,  $\Phi \approx 3.32$ 
and $\xi \approx 38$ for the enhanced case and 
$\alpha \approx 2.0$, $C \approx 136$, 
$\beta \approx 1.13$,  $\Phi \approx 1.78$ 
and $\xi \approx -87$ for the quenched case.
It is not clear if $n>3$ is enough to reach outside the preasymptotic region and
the values may have to be taken as qualitative asymptotic results. 
Nevertheless, there is
a big difference in $\alpha$ which indicates that the coupling is of  
longer range in the enhanced case.
There is also a phase shift 
analogous to the magnetization results.
Additionally, the strength of the FM 
peaks at 4 ML and 11 ML is strongly reduced in the quenched case. 
The differences  must
be attributed to the exchange enhancement in the Pd spacer which thus has large
qualitative as well as quantitative effect on the MIC.

\begin{figure}
    \includegraphics*[angle=0,width=0.45\textwidth]{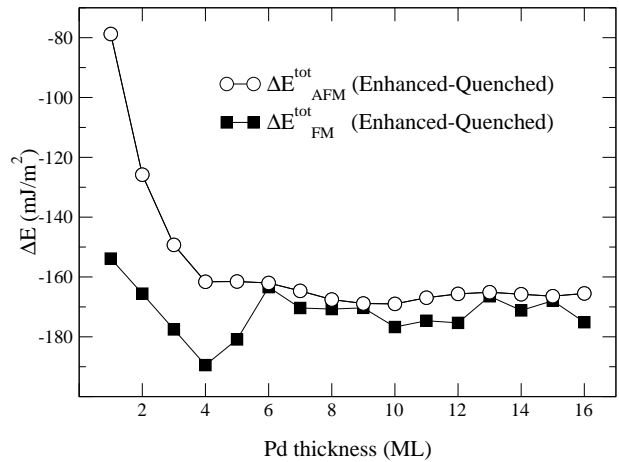}
    \caption[]{The energy gain due to the  exchange enhancement in the FM and AFM 
                    configurations. The only case that contributes to the strong  
                    positive peaks in the MIC is the FM configuration. 
                    \label{fig:EAEF}}
\end{figure}

In order to determine for which configuration the enhancement has the 
largest effect we have calculated the difference in total energy between the 
enhanced (e) and the quenched (q) calculations for the AFM and FM systems 
separately. The formula $\Delta E^{tot}_{A(F)} = 
E^{e}_{A(F)}-E^{q}_{A(F)}$ was used where $A$ and $F$ denotes the 
antiferromagnetic and ferromagnetic configurations, respectively. The result is displayed in 
Fig.(\ref{fig:EAEF}) and
it is evident that the system always 
lowers its energy when the
quenching is removed. The lowering of the energy 
is almost the same for all thicknesses above 4 ML in the AFM case but 
shows a similar oscillating structure as the enhanced MIC for  the 
FM case. One can also see that the system always lowers its energy
more in the FM configuration than in the AFM.
From this observation we may draw the conclusion that the 
strong FM peaks of the MIC as well as the FM bias originates from 
the magnetic enhancement in the FM configured system. 
Note that it is very
unlikely to be an effect of changed Fe magnetization, since
we find almost the same moment for the Fe in the enhanced and quenched 
calculations (see Table \ref{tab:mprofile}). We have also shown that the 
two cases has very similar reflection coefficients so this possibility
is also excluded.

The energy gain by splitting the bands is of course larger if the 
Fe layers at the interfaces have high magnetic moments and thus induces high 
local moments on the Pd atoms. Then by reducing the magnetic 
moments at the Fe interfaces, for example by growing a few more fcc layers
or by growing a thicker Fe film in the bcc structure, 
it may be possible to reduce the FM
bias and thus obtain a weak AFM coupling in the region 
$6\le N \le9$. 
The MIC of the quenched case should be well described by 
conventional RKKY/QW models. However, the drastic difference between these
calculations and the full calculation imply that the latter is not described 
by these models.

\section{Summary}
We have investigated the magnetic properties of Fe/Pd fcc (100) systems and found 
a giant magnetic enhancement (GME) effect on the induced moments. 
We have estimated the effect of the GME on the MIC for  Fe/Pd/Fe bilayer systems 
and concluded that the paramagnetic groundstate of the spacer is modified to 
such an extent that perturbation theories like RKKY  and the QW model 
for the MIC may not be straightforward to apply. In particular, the overall FM bias, 
the asymptotic decay and the phase shift may not be captured. 
In the same bilayer system we have also 
found very strong QW induced magnetic enhancement on the same order of 
magnitude as the GME effect. 
We suggest that the strength of the magnetic moments at the interfaces in the 
Fe/Pd/Fe system may determine if an AFM coupling can be observed.

\begin{acknowledgments}
One of us (E.H.) like to thank prof. Balazs Gy{\"o}rffy for great hospitality and valuable
discussions. The help from Olle Eriksson and Igor Abrikosov is also acknowledged.
This work was supported from the Swedish Research Council (VR), the Swedish Foundation for
Strategic Research (SSF) and the European Network for Computational 
Magnetoelectronics. The work by (A.M.N.N.) was performed under the auspices of
The U.S. Department of Energy.
\end{acknowledgments}

\end{document}